\documentclass[preprint,preprintnumbers,amsmath,amssymb,show keys,showpacs,secnumarabic]{revtex4}

\usepackage[dvips]{graphicx}
\usepackage{amsmath,amsthm,amssymb}
\usepackage[dvips]{color}

\begin{document}

\title{Characterization of foreign exchange market\\using the threshold-dealer-model}%

\author{Kenta Yamada$^1$}\email[E-mail: ]{yamada@smp.dis.titech.ac.jp}
\author{Hideki Takayasu$^2$}
\author{Misako Takayasu$^1$}
\affiliation{$^1$Department of Computational Intelligence and Systems Science, Interdisciplinary Graduate School of Science and Engineering, Tokyo Institute of Technology, 4259 Nagatsuta-cho, Midori-ku, Yokohama 226-8502, Japan}

\affiliation{$^2$Sony Computer Science Laboratories, 3-14-13 Higashi-Gotanda, Shinagawa-ku, Tokyo 141-0022, Japan}

\begin{abstract}
We introduce a deterministic dealer model which implements most of the empirical laws, such as fat tails in the price change distributions, long term memory of volatility and non-Poissonian intervals. We also clarify the causality between microscopic dealers' dynamics and macroscopic market's empirical laws.
\end{abstract}

\maketitle
\section{introduction}
Mathematical models of open markets can be categorized into two types. In one type, the market price time series are directly modeled by formulation such as a random walk model, ARCH and GARCH models\cite{ARCH}\cite{GARCH}, and the potential model\cite{potentialmodel}\cite{potentialmodel2}\cite{potentialmodel3}. The other type is the agent-based model which creates an artificial market by computer programs\cite{dealermodel}\cite{izumi}\cite{lux}. The agent-based model is able to clarify the relationship between dealers' actions and market price properties. Just like the simple ideal gas model reproducing basic properties of real gas, we can expect that simple dealers' actions can reproduce the empirical laws of market prices. 

     In this paper we systematically introduce 4 deterministic dealer models in order to clarify the minimal actions of dealers to satisfy the empirical laws of markets. These are revised models of so-called the threshold model which is originally introduced by one of the authors (H.T) and coworkers\cite{dealermodel1} in order to demonstrate that dealers' simple actions can cause deterministic chaos resulting the market price apparently random even the dealers' actions are completely deterministic. We revise the model step by step to reconstruct most of the empirical laws. 

\section{Construction of the dealer model}
\subsection{The model-1}
We start with the model-1, the simplest model. In this model, dealers' dynamics is defined by the following differential equation;
\begin{equation}
\frac{db_i}{dt^{\prime}}=\sigma_i c_i
\end{equation}
where $\sigma_i$ indicates

\begin{eqnarray}
\sigma_i=\left\{
\begin{array}{l}
{+1}\quad \text{buyer}\nonumber\\
{-1}\quad \text{seller}\nonumber
\end{array}
\right.
\end{eqnarray} 

In the dealer model, the $i$-th dealer offers the limit price which consists of bid price($b_i$) and ask price($a_i$). For simplicity we assume that the value of spread is a constant($L$), so the ask price($a_i$) is given by $a_i=b_i+L$. If a dealer wants to buy($\sigma_i=+1$) but the transaction was unsuccessful, he will raise his price at a given rate $c_i$ until he can actually buy. For a seller($\sigma_i=-1$) he will reduce his price until he completes a successful transaction. Large $c_i$ means that the $i$-th dealer is quick-tempered. The values of $c_i$ are given randomly as an initial condition representing each one's disposition is different. This heterogeneity is important for realizing trades. A trade takes place if the model's parameters satisfy,
\begin{equation}
\max\{b_i(t)\}-\min\{a_j(t)\}\ge 0
\end{equation}
The market price is quoted between bid and ask prices. We assume that trading dealers have a minimum unit of yen or dollar. After a trade the seller and the buyer change places. For example, assume the $i$-th dealer; he is a buyer($\sigma_i =+1$) and after making a purchase, becomes a seller($\sigma_i =-1$).

Although the dynamics is completely deterministic, apparently the market prices random walk. By the effect of transactions the limit prices are attracted to each other if the distance of any two pair of prices exceeds the spread, L. This effect causes negative values of autocorrelation on the first few ticks(FIG. 1-B). Both the autocorrelations of volatility and transaction interval decay quickly(FIG. 1-C,D). Volatility is defined by the absolute values of price changes. No volatility clustering can be observed in this model(FIG. 1-A) and the occurrence of transaction is fairly modeled by a Poisson process. In fact, the model-1 can be understood as a uniform random noise generator. However, the real market data are characterized by fat tails in the price change distributions, long volatility correlations and Non-Poissonian intervals. So, we add a new effect each to the model-2 and model-3.
\begin{figure}[htbp]
	\begin{center}
	    \includegraphics[width=8.0cm]{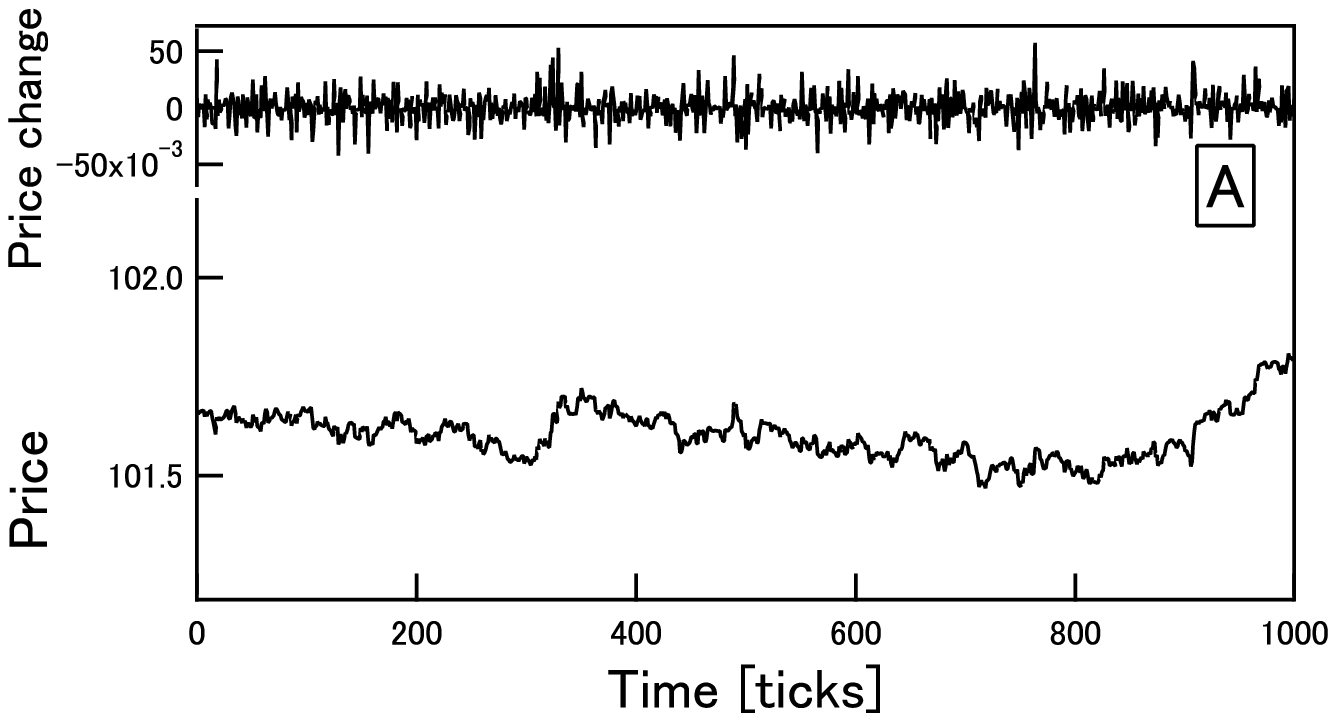}
	    \includegraphics[width=8.0cm]{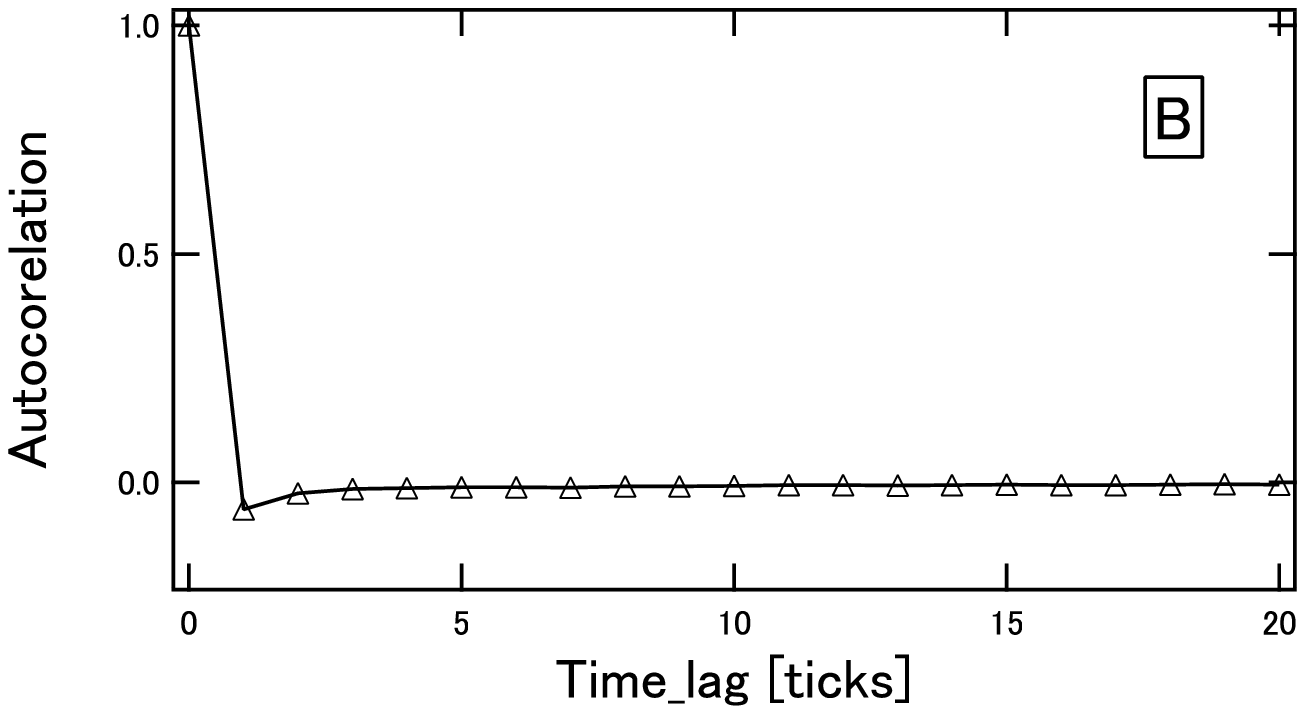}
	    \includegraphics[width=8.0cm]{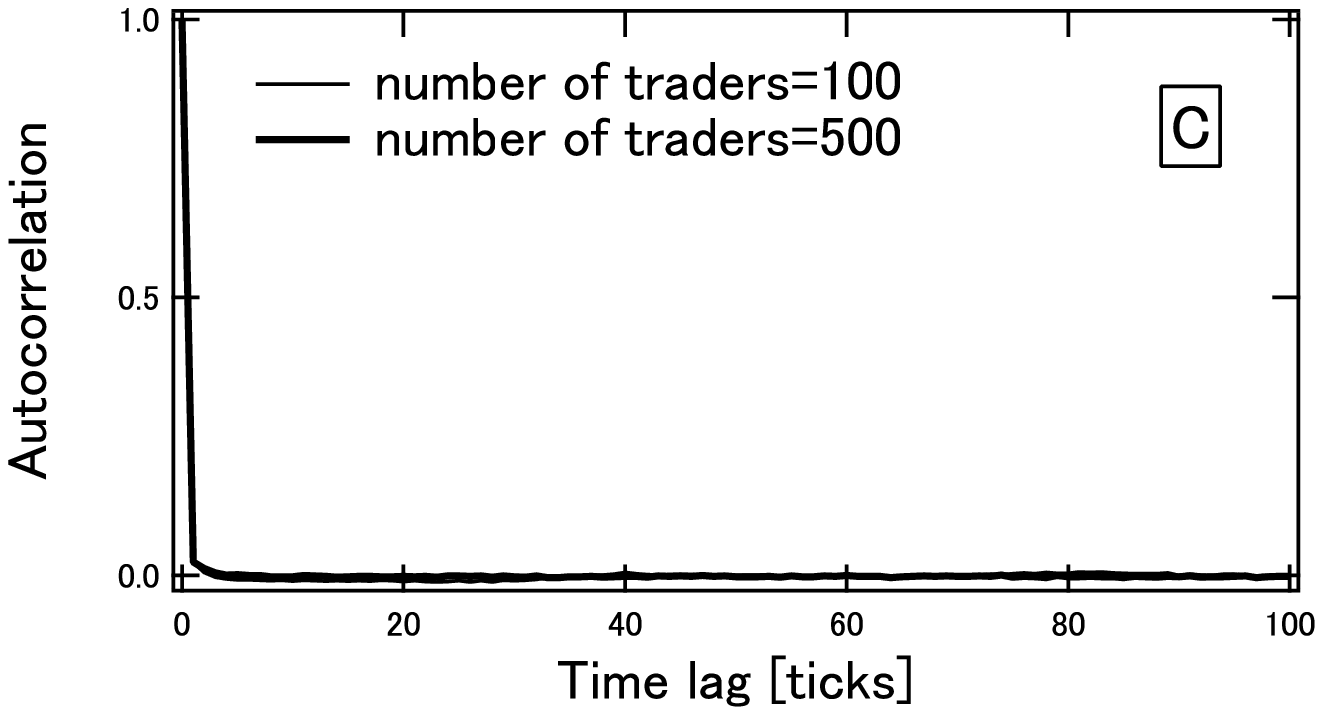}
	    \includegraphics[width=8.0cm]{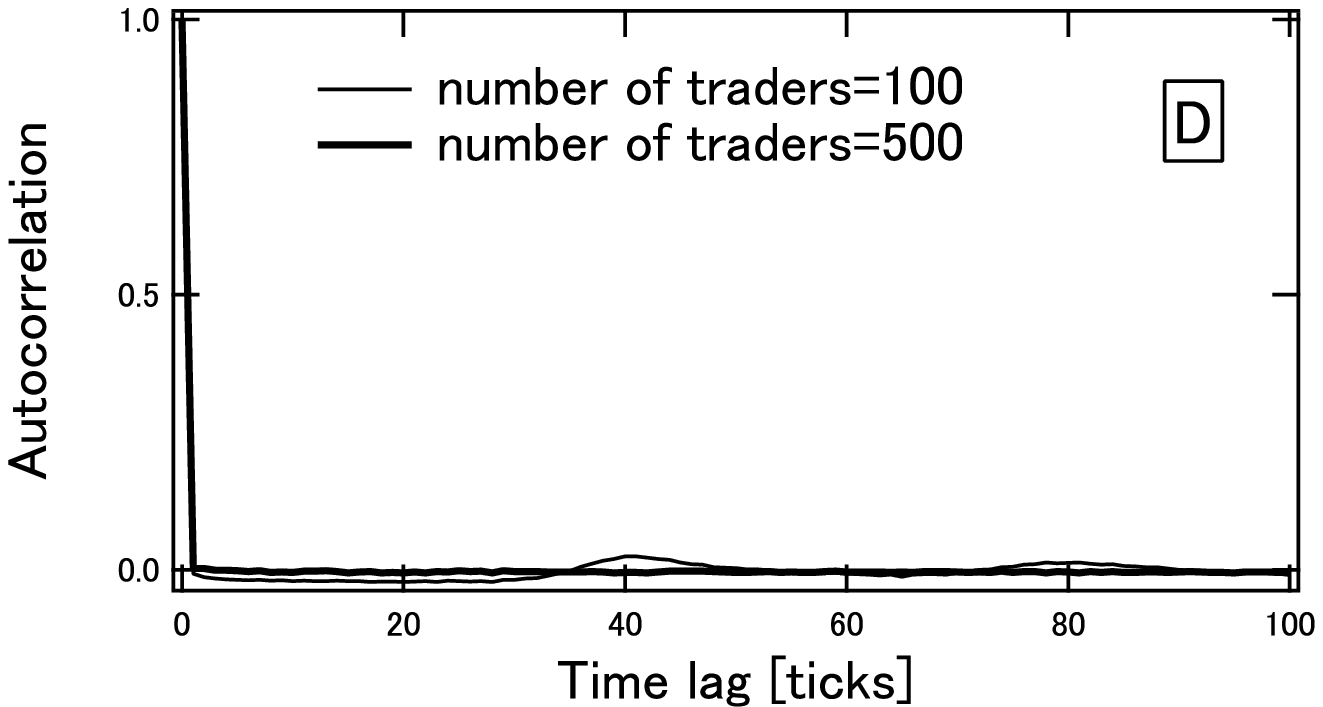}
	\end{center}
    \caption{Results of the model-1. We set parameters: $N$(number of dealers)=100, $L$(spread)=1.0, $dt$=0.01, $c_i$=[-0.015,0.015].}
\label{interval}
\end{figure}
\newpage

\subsection{The model-2}
In the model-2 we focus on transaction intervals. In order to make our model closer to the real market we add so-called the "self-modulation" effect to the model-1\cite{transaction interval}. An example of the "self-modulation" effect can be found in FIG. 2 of real data. We use a set of tick-by-tick data for the Yen-Dollar exchange rates for the 3 month from January 1999 to March 1999. We pay attention to the periods where the frequency of trading is high, such as 6:00$\sim$11:00 on New York time. The time intervals of transactions tend to make clusters, namely, shorter intervals tend to follow after shorter intervals, and longer intervals follow longer intervals. The distribution of transaction time intervals has longer tail than exponential distribution of Poisson process. It is known that this effect is well-modeled by the "self-modulation" process\cite{transaction interval} and that normalized transaction interval by moving averaged transaction interval becomes white noise. In the model-1 we can find $F_k=T^{\prime\alpha}_k/\overline{T^{\prime}}$ ($\alpha\sim 1.25$) has same statistical properties to the normalized transaction interval in the real market. $T^{\prime}_k$ is transaction interval between (k-1)-th and k-th ticks in the model-1 and $\overline{T^{\prime}}$ is average of $T_k^{\prime\alpha}$ .  So we modulate this noise and we can reproduce the transaction interval having the same properties to transaction interval in the real market. Mathematical expression is given by     

\begin{equation}
\begin{cases}
& {\displaystyle\frac{db_i}{dt}}={\displaystyle\frac{1}{G_k<T_k>_{\tau}}}\sigma_i c_i\label{eq:model1}\\
& \\
& dt=G_k<T_k>_{\tau}dt^{\prime}
\end{cases}
\end{equation}
   Here, $G_k=T^{\prime\beta}_k/\overline{T^{\prime}}$ ($\beta= \alpha-1.0$). The model-2 has two time axes, $t^{\prime}$ and $t$. $t^{\prime}$ is the same time axes of the model-1. On the other hand $t$ takes into account the "self-modulation" effect. $t^{\prime}$ is the normal time step in the simulation and $t$ corresponds to real time. $<T_k>_\tau$ is a moving avarage of $T_k$ which is for last $\tau$ seconds. So, the term of $dt$ in Eq.(3) depends on $<T_k>_\tau$, in other words $dt$ relies on the past transaction intervals. This is "self-modulation" process in mathematical expression.
    
By adding this effect random intervals of the model-1 are modulated and the distribution of intervals has fat tails(FIG. 3), thus becoming closer to the real data.
\begin{figure}[htbp]
	\begin{center}
	    \includegraphics[width=8.0cm]{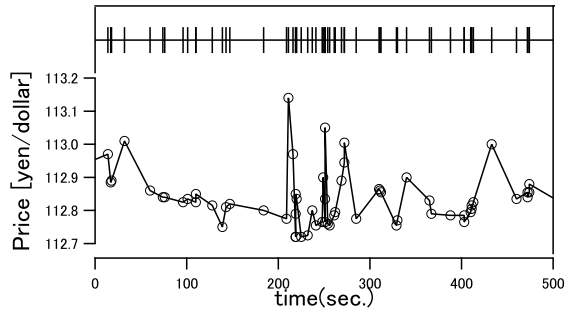}
	\end{center}
    \caption{Time series of real yen/dollar rate and the transaction intervals.}
\label{interval}
\end{figure}
\begin{figure}[htbp]
	\begin{center}
	    \includegraphics[width=8.0cm]{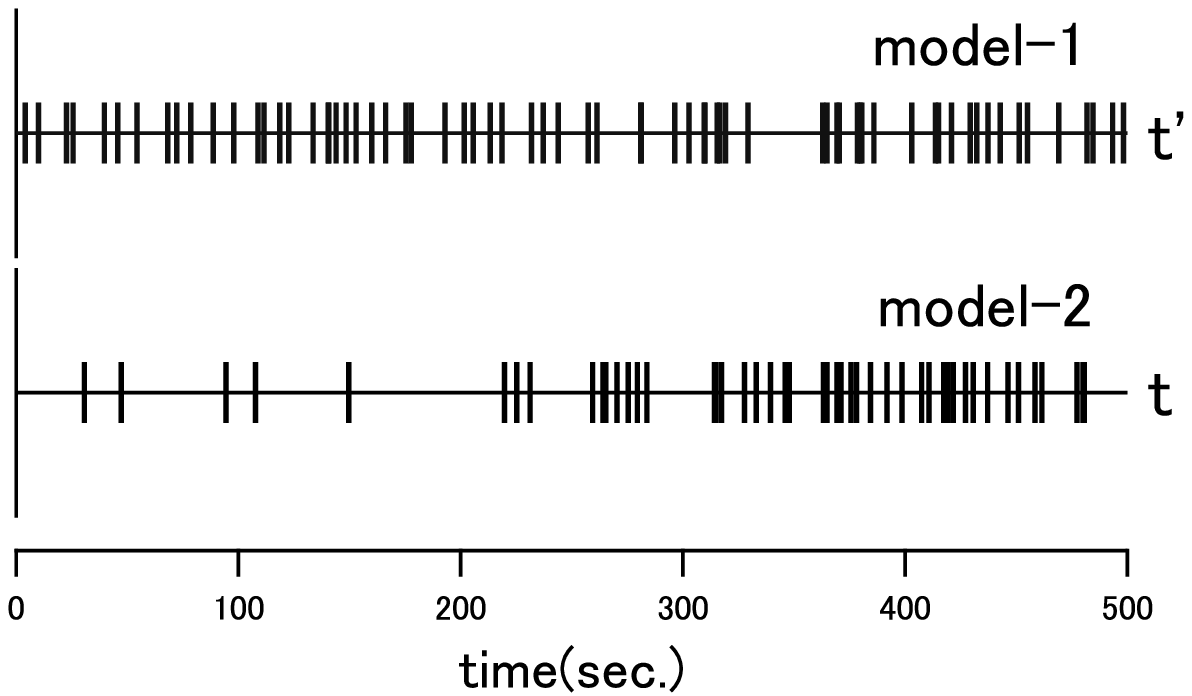}
	    \includegraphics[width=8.0cm]{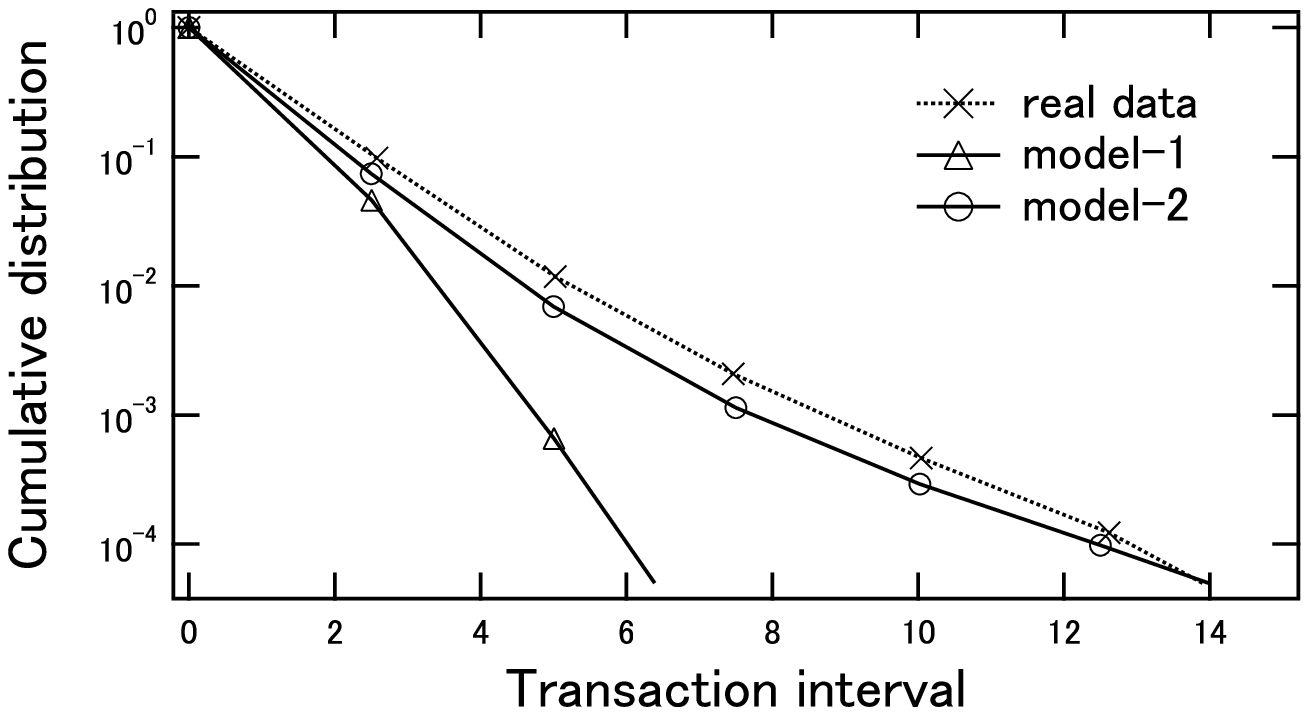}
	\end{center}
    \caption{Comparison of transaction intervals between the model-1 and model-2. The left figure represents transaction intervals like Fig.2, and the figure in the right hand side cumulative distribution of transaction intervals. In this figure we normalize transaction intervals by average. In the model-2 we use the parameters, $N$(number of dealers)=500, $L$(spread)=1.0, $dt$=0.01, $c_i$=[-0.01,0.01] and $\tau$=150.}
\label{interval}
\end{figure}
\newpage

\subsection{The model-3}
In the model-3 we cast a spotlight on the volatility. In the model-1 the distribution of volatility is exponential as shown in FIG.4. Now we add a new effect to the model-1, namely the feedback effect of price changes to the dealers. "Trend-follow" is a tendency of dealers' action that they predict price rise when the market price has risen. Mathematically we add a term that is proportional to the latest averaged price change, $<dP>_{\eta}$.
\begin{equation}
\frac{db_i}{dt^{\prime}}=\sigma_i c_i+\epsilon_i d_i<dP>_{\eta}\label{eq:model3}
\end{equation}
Here $\epsilon_i$ and $<dP>$ are

\begin{eqnarray}
\epsilon_i=\left\{
\begin{array}{l}
{+1}\quad \text{follower}\nonumber\\
{-1}\quad \text{contrarian}\nonumber
\end{array}
\right.
\end{eqnarray}

\[
<dP>_{\eta}=\sum_{t=0}^{\eta} \frac{w_j}{W} dP_t
\]
The term $w_j$ is the $j$-th weight and $W$ is normalization factor that is, $W=\sum w_j$.
Here, we assume that dealers' response to the trends are different, and initially the coefficients $d_i$ are randomly assigned. This causes the tails of distribution of volatility being stretched(FIG. 4). The functional form of this distribution depends on the values of $d_i$\cite{dealermodel2}.

\begin{figure}[htbp]
	\begin{center}
	    \includegraphics[width=8.0cm]{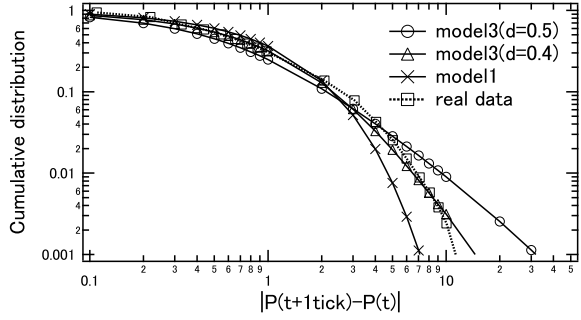}
	\end{center}
    \caption{Cumulative distribution of volatility. The values are normalized by average. We can observe that the tails are shifted by the "trend-follow" effect. In the model-3 we use the parameters, $N$(number of dealers)=100, $L$(spread)=1.0, $dt$=0.01, $c_i$=[-0.015,0.015], the value of $d_i$ is expressed in the figure, and $<dP>$ is the latest price change.}
\label{interval}
\end{figure}

\subsection{The model-4}
Finally, we combine the model-2 and model-3 to make the model-4. Mathematical expression is given as follows

\begin{equation}
\begin{cases}
& {\displaystyle\frac{db_i}{dt}}={\displaystyle\frac{1}{G_k<T_k>_{\tau}}}\left(\sigma_i c_i+\epsilon_i d_i{<dP>_{\eta}}\right)\\
& \\
& dt=G_k<T_k>_{\tau}dt^{\prime}
\end{cases}
\end{equation}
In the model-4 we set $w_j=\exp(-0.3j)$ and $\eta=150$. Hence we know that the dealer refers to the last 150 seconds and the weights decay exponetially\cite{MA_weight}. As expected we can reproduce most of the empirical laws established in real markets. The autocorrelation of price change decays quickly after a negative value at 1 tick(FIG. 5-C). We can observe that volatility clearly clusters(FIG. 5-B), and that the autocorrelation of volatility has a long tail similar to the real data(FIG. 5-D). Also, the distribution of transaction intervals and volatility are both reasonably close to the real market data(FIG. 5-E,F). Moreover, the diffusion property, the standard deviation as a function of time, is fairly close to that of real data(FIG. 5-G).
\begin{figure}[htbp]
	\begin{center}
	    \includegraphics[width=8.0cm]{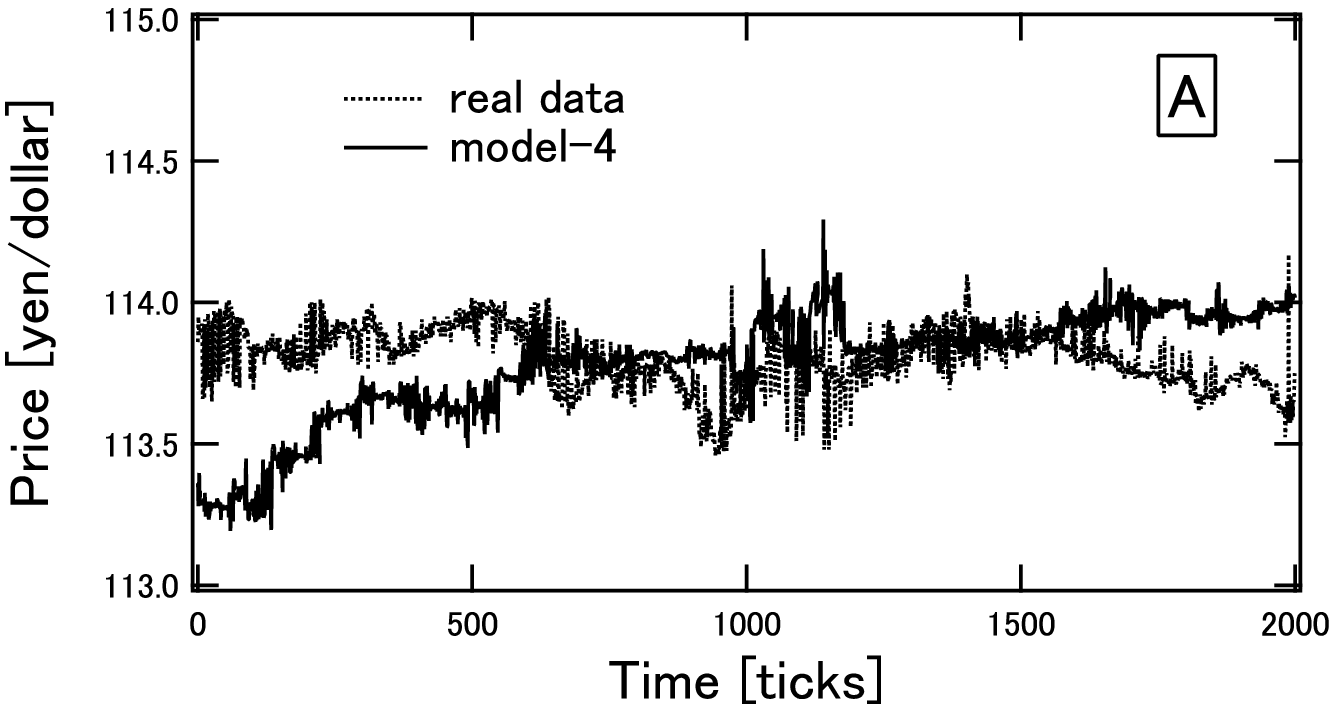}
	    \includegraphics[width=8.0cm]{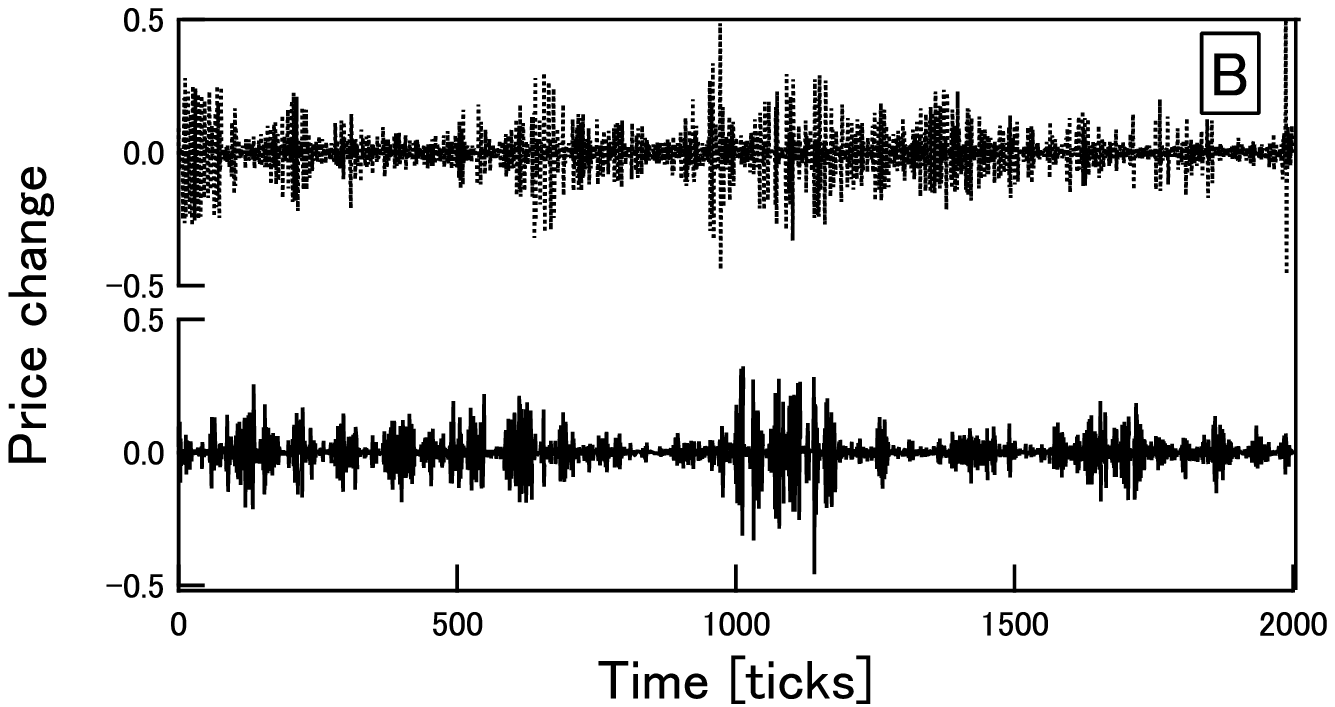}
	    \includegraphics[width=8.0cm]{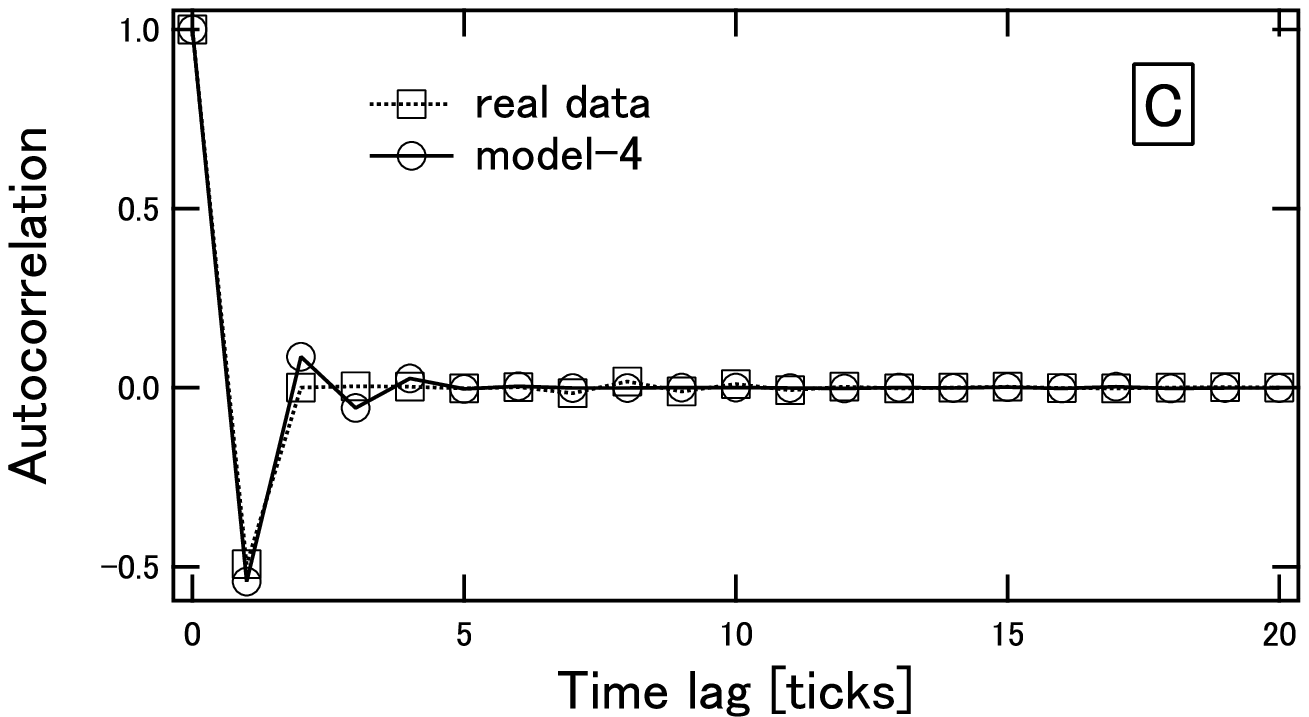}
	    \includegraphics[width=8.0cm]{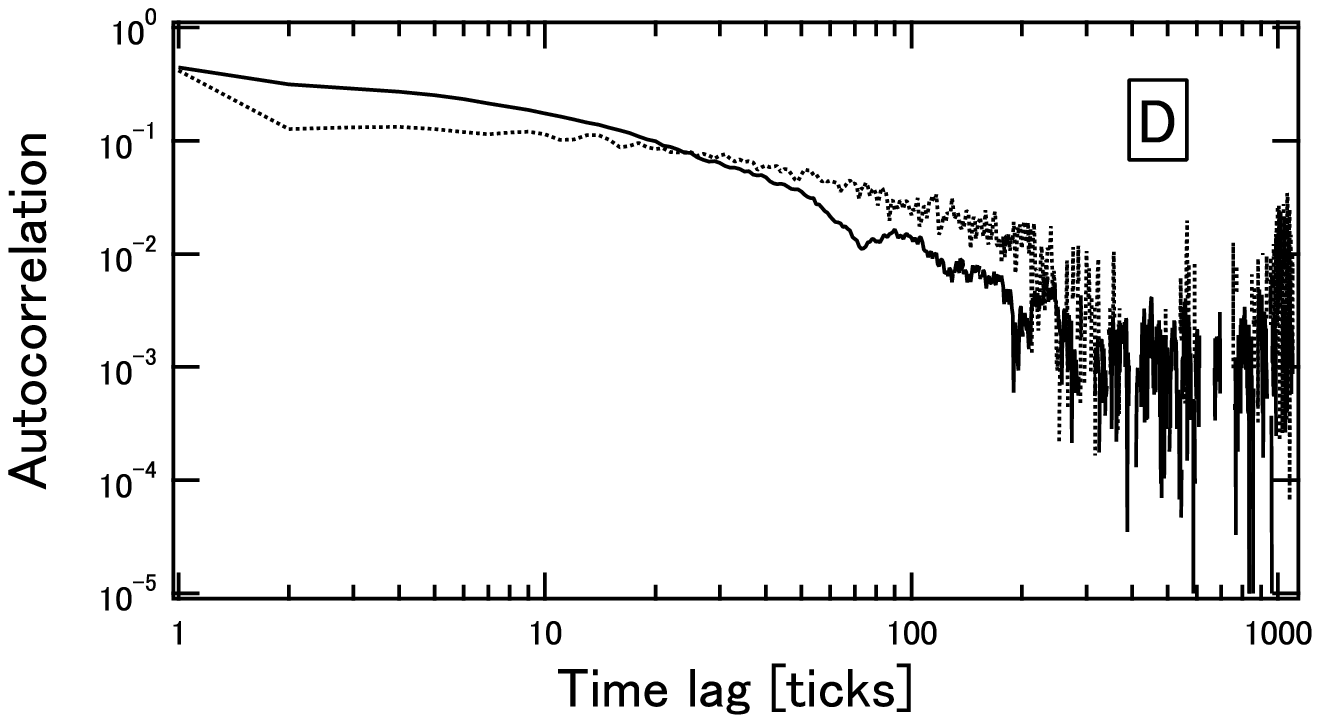}
	    \includegraphics[width=8.0cm]{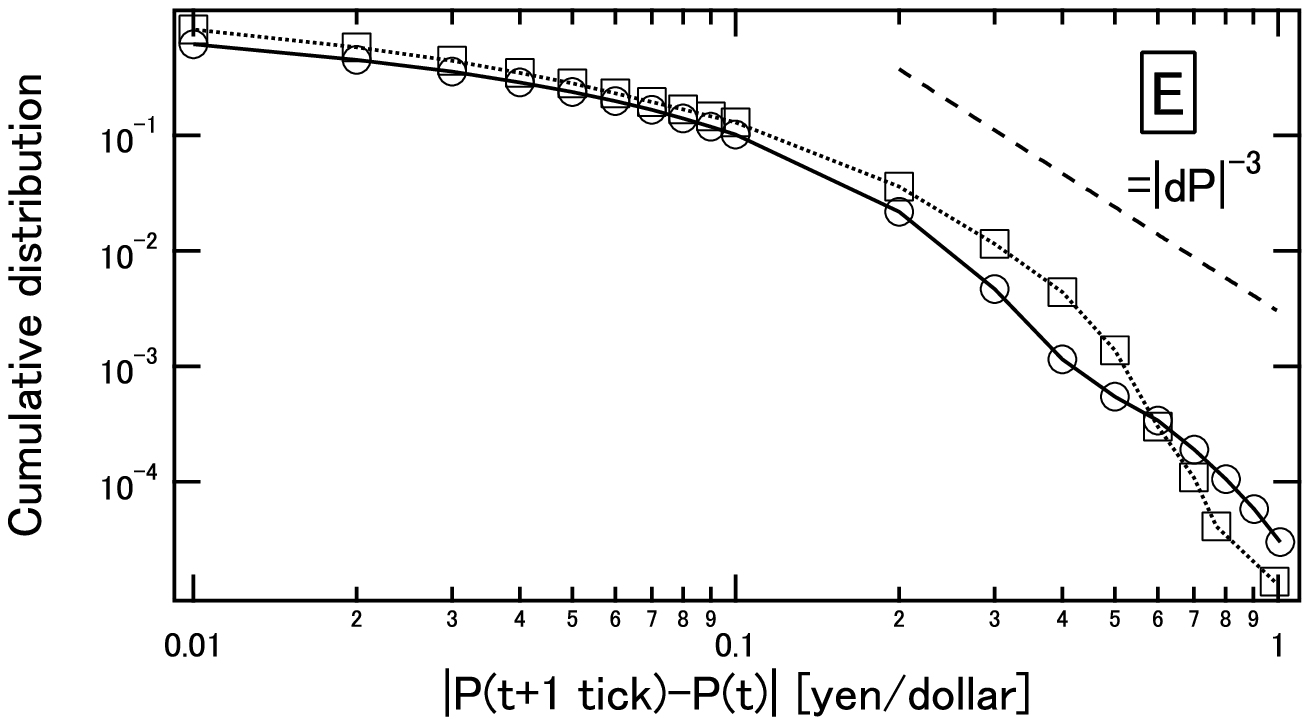}
	    \includegraphics[width=8.0cm]{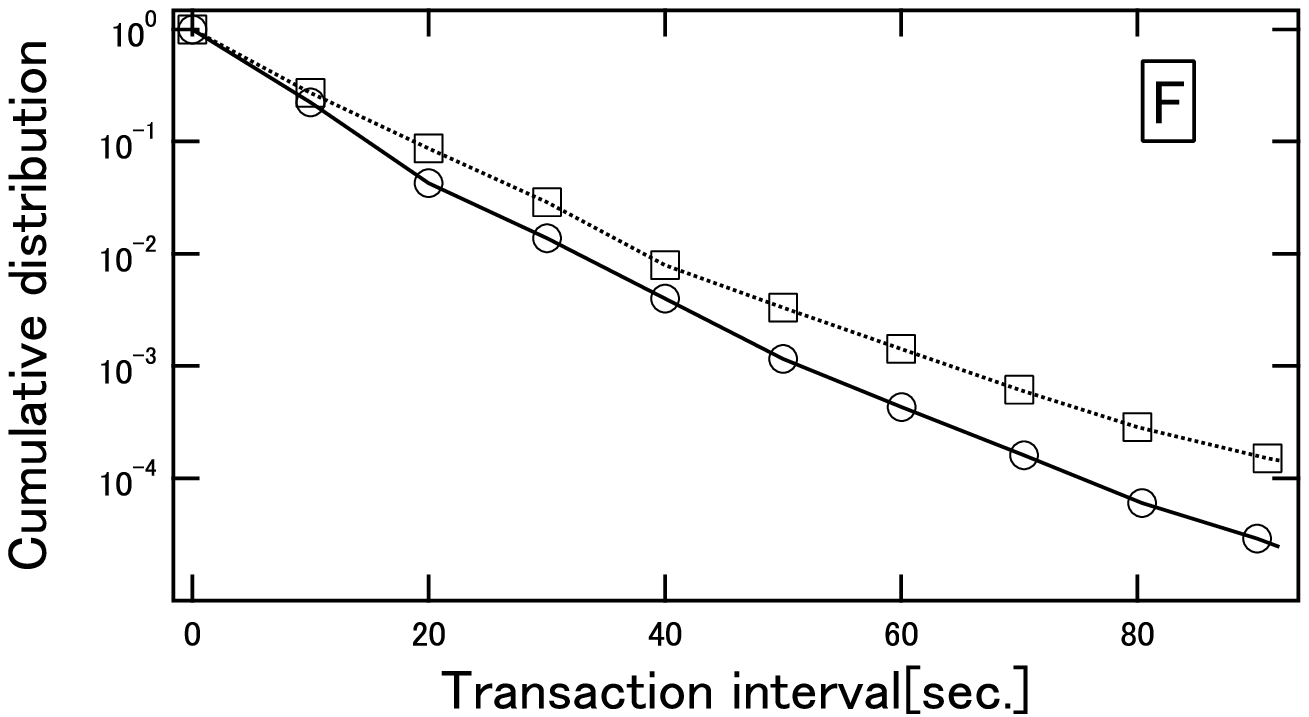}
        \includegraphics[width=8.0cm]{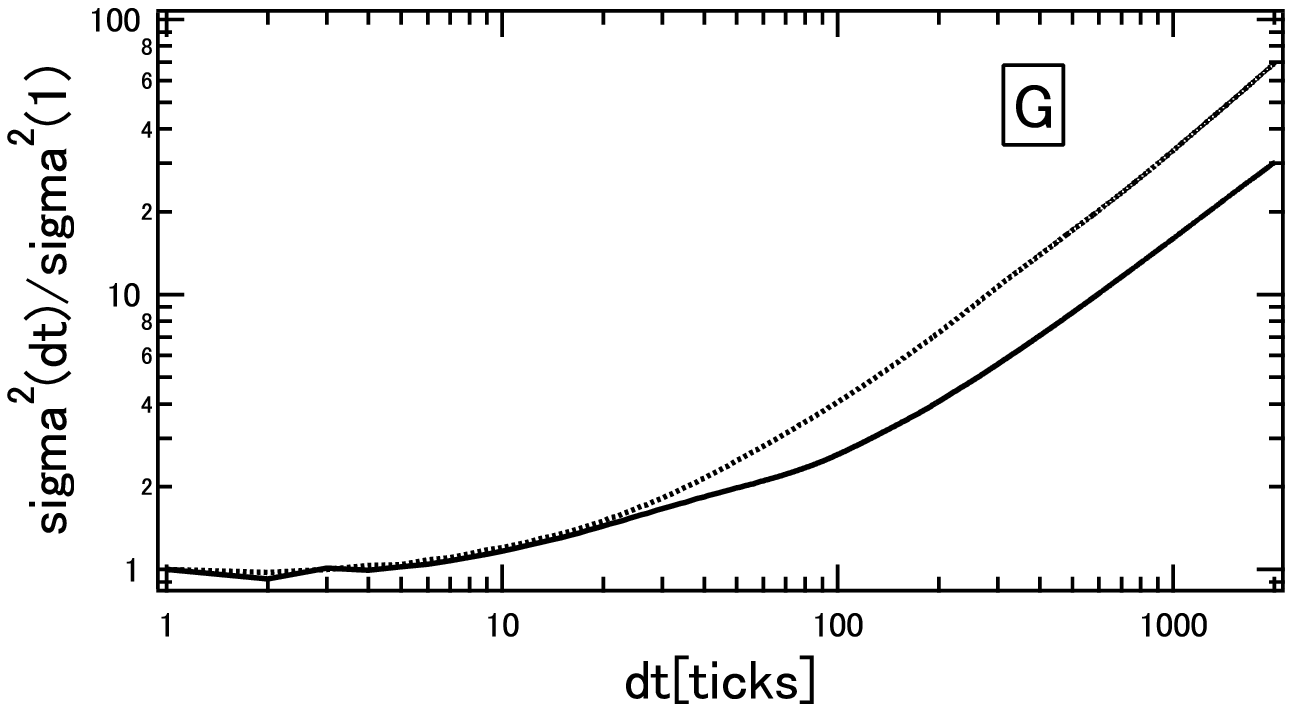}
	\end{center}
    \caption{Results of the model-4. The values of the model-4 are calculated by using the parameters, $N$(number of dealers)=100, $L$(spread)=1.0, $dt$=1.0, $c_i=[-0.02,0.02]$, $\tau=150$, $\eta=150$ and $d_i=[5.0,7.0]$ }
\label{interval}
\end{figure}

\newpage

\section{conclusion}
We started with the model-1, a very simple model. Considering why the model-1 differs from real market data, we added two effects to the model-1, which are feedback effects. One of them is the "self-modulation" used in the model-2, and the other is the "trend-follow" used in the model-3. The model-4, applying both of the effects, satisfies most of the empirical laws. It should be noted that each dealer has only three parameters describing his character. Finally, we summarize our results in TABLE \ref{tb:1}
. 

Sato and Takayasu already showed that the dealer model's price fluctuations can be approximated by ARCH model in some conditions\cite{dealermodel-ARCH}. Such approach of connecting the one type of market models to another will be fruitful study in the near future.
   
\begin{table}[htbp]
\caption{Results of each model.}
 \begin{center}
  \begin{tabular}{|l|c|c|c|c|}
    \hline
       &  model-1  & model-2   & model-3   & model-4   \\
    \hline
   Correlation of price change    & B   & B   & B   & A   \\
    \hline
   Distribution of volatility    &  $-$  & $-$   & A   & A   \\
    \hline
   Correlation of volatility    &  $-$  & $-$   & $-$   & A   \\
    \hline
   Distribution of intervals    &  $-$  & A   & $-$   & B   \\
    \hline
   Diffusion of price    & $-$   & $-$   & $-$   & B   \\
    \hline
    \multicolumn{5}{c}{A:satisfy(quantitatively) B:satisfy(qualitatively) $-$:not satisfy}\\ 
   \end{tabular}
 \end{center}
 \label{tb:1}
\end{table}

\begin{acknowledgements}
This work is partly supported by Japan Society for the Promotion of 
Science, Grant-in-Aid for Scientific Research $\sharp 16540346$ (M.T.) .
\end{acknowledgements}

\end{document}